\documentstyle[preprint,revtex,eqsecnum]{aps}

\begin{document}
\preprint{PITT 92-06 and LPTHE-PAR 92--33}

\vspace{3mm}

\draft
\begin{title}
{\bf Quantum Rolling Down Out of Equilibrium}
\end{title}
\author{{\bf D. Boyanovsky}}
\begin{instit}
Department of Physics and Astronomy\\ University of
Pittsburgh \\
Pittsburgh, P. A. 15260, U.S.A.
\end{instit}
\author{\bf H. J. de Vega}
\begin{instit}
Laboratoire de Physique Theorique et Hautes
Energies\cite{lpthe}\\Universite
Paris VI\\Tour 16, 1er etage, 4, Place
Jussieu\\75252 Paris cedex 05, FRANCE.
\end{instit}
\begin{abstract}
In a scalar field theory, when the tree level potential
admits broken symmetry ground states, the quantum
corrections to
the static effective potential are complex. (The imaginary
part
is a consequence of an instability towards phase separation
and
the static effective potential is not a relevant quantity
for
understanding the dynamics). Instead, we study here the  equations of
motion obtained from the one loop effective action for slow rollover
out of equilibrium.
 We considering the
case in which a scalar field theory undergoes a rapid phase
transition from $T_i>T_c$ to $T_f<T_c$. We find that, for
slow rollover
initial conditions (the field near the maximum of the tree
level potential), the process of
phase separation controlled by unstable long-wavelength
fluctuations
introduces
dramatic corrections to the dynamical evolution of the
field.
We find that these effects  slow the rollover even further
thus delaying the phase transition, and increasing the time
that the field spends near the ''false vacuum''. Moreover,
when the initial value of the field is very close to zero,
the dynamics becomes {\it non-perturbative}.
\end{abstract}
\pacs{}

\newpage

\section{\bf Introduction and Motivation}

Inflationary cosmological models provide very appealing
scenarios
to describe the early evolution of our
universe\cite{abbott}.
Since the original model proposed by Guth\cite{guth1},
several
alternative scenarios have been proposed to overcome some of
the
difficulties with the original proposal.

Among them, the new inflationary
model\cite{guth2,linde1,steinhardt,linde2}
is perhaps one of the most attractive. The essential
ingredient
in the new inflationary model is a scalar field that
undergoes a
second order phase transition from a high temperature
symmetric phase
to a low temperature broken symmetry phase. The expectation
value (or
thermal average) of the scalar field $\phi$ serves as the
order parameter.
Initially at high temperatures, the scalar field is assumed
to be in
thermal equilibrium and  $\phi \approx 0$.  The usual field-
theoretic tool
to study the phase transition is the effective
potential\cite{kirzhnits,dolan,sweinberg}. At high
temperatures,
the global minimum of the effective
potential is at $\phi =0$, whereas at low temperatures there
are two
degenerate minima.

The behavior of the phase transition in the new inflationary
model is the
following: as the universe cools down the expectation value
of the scalar
field
remains close to zero until the temperature becomes smaller
than the
critical temperature. Below the critical temperature, when
the effective
potential develops degenerate minima away from the origin,
the scalar
field begins to ``roll down the potential hill''. In the new
inflationary
scenario, the effective potential below the critical
temperature is
extremely flat near the maximum, and the scalar field
remains near the
origin, i.e. the false vacuum for a very long
time and eventually rolls down the hill very slowly.
This scenario thus receives  the name of ``slow rollover''.
During the slow rollover stage, the energy density of
the universe is dominated by the constant vacuum energy
density $V_{eff}(
\phi=0)$, and the universe evolves rapidly into a de Sitter
space (see for
example the reviews by Kolb and Turner\cite{kolb}, Linde\cite{linde}
and Brandenberger\cite{brandenberger}). Perhaps the most remarkable
consequence of the new
inflationary scenario and the slow rollover transition is
that they provide a calculational framework for the prediction
of density fluctuations\cite{starobinsky}. The coupling
constant in the
typical zero temperature potentials must be fine tuned to a
very small value to reproduce the observed limits on  density
fluctuations\cite{kolb,linde}.

This picture of the slow rollover scenario, is based on the
{\it static}
effective potential. The use of the static effective
potential to describe
a time dependent situation has been criticized by Mazenko,
Unruh and
Wald\cite{mazenko}. These authors argued that the {\it
dynamics} of the
cooling down process is very similar to the process of phase
separation in
statistical mechanics. They argued that the system will form
domains and that the scalar field will relax to the values
at the minima of
the potential very quickly.

Guth and Pi\cite{guthpi}, performed a thorough analysis of
the effects
of quantum fluctuations on the time evolution. These authors
analyzed
the situation below the critical temperature by treating the
potential
near the origin as an {\it inverted harmonic oscillator}.
They recognized that the
instabilities associated with these upside-down oscillators
lead to an
exponential growth of the quantum fluctuations at long times
and to a
classical description of the probability distribution
function. Guth
and Pi also recognized that the {\it static} effective
potential is not
appropriate to describe the dynamics, that must be treated
as a time
dependent process.

Subsequently, Weinberg and Wu\cite{weinbergwu}, have studied
the effective
potential, particularly in the situation when the tree level
potential
allows for broken symmetry ground states. These authors
carefully analyzed
the contributions to the effective potential and showed that
the imaginary
part of the one-loop effective potential is the result of an
analytic continuation of the unstable modes (inverted oscillators)
studied by  Guth and Pi. This imaginary part is in fact concealing the
growth of this unstable modes and a time dependent situation that cannot be
described by equilibrium statistical mechanics. The imaginary part of the
effective potential was related to the lifetime of a particular
initial quantum state.

 There have been several attempts to study the time
evolution of the scalar
 field either in flat spacetime or de Sitter
 space\cite{ringwald,kripfganz,leutwyler,eboli,samiulla},
but to our
 knowledge, the influence of the instabilities that are
responsible for
 domain growth and phase separation, on the dynamics of the
scalar field
 has not  yet been elucidated.

 In this article we study the quantum dynamics of the scalar
field by
 analyzing
 the situation in which the system originally in equilibrium
at a
 temperature higher that the critical temperature evolves
{\it out of
 equilibrium} at a temperature below the critical
temperature. We provide
 a detailed description of the dynamics out of equilibrium,
concentrating
 on the instabilities that drive the phase transition and
phase separation,
 that is the unstable growth of long-wavelength
fluctuations.
 Our approach consists in obtaining the non-equilibrium
evolution equation
 obtained from the one-loop effective action. These
equations turn out to be
 non-local and non-integrable\cite{integrable}. We provide a
qualitative discussion and a consistent numerical analysis of these
evolution equations for a wide variety of initial conditions.

 The results of our analysis show that the instabilities
that trigger the
 growth of long-wavelength fluctuations dramatically enhance
the quantum corrections and result in a further slow-down of the scalar
field.

 For ``slow-rollover'' initial conditions,
 we show that even for very weak coupling (consistent with
the bounds on density fluctuations), the quantum corrections
become very  important, and slow the dynamics dramatically, and in
particular for some  initial conditions, in which the scalar field
is very close to the "false
vacuum", the quantum corrections must be treated beyond
perturbation theory.

 This paper is organized as follows: in the next section we
review the
 effective potential and its shortcomings to describe the
dynamics.

In section (III) we describe the non-equilibrium formalism
and our
approach to obtain the  equations of motion out of
equilibrium for the spatial average of the scalar
field. In this section, we provide an
analysis of the renormalization aspects of the effective
equations of motion.

In section (IV) we study the dynamics and provide an
analytic as well as
numerical analysis of the evolution equations for a wide
range of initial
conditions emphasizing the consequences of  the unstable
growth of  long wavelength fluctuations.

We conclude the paper with an analysis of the potential
implications
of our results on cosmological phase transitions and
inflationary models.

\section{\bf The Complex Effective Potential}

It is already well known that when the tree level potential
allows for broken symmetry ground states, the one loop
effective
potential becomes complex in a region of field
configurations.
This unacceptable complex part of the effective potential
has
been deemed an artifact of the loop expansion, and for the
most
part ignored in most treatments of the effective potential.

However as was clearly shown by Weinberg and
Wu\cite{weinbergwu},
the imaginary part of the effective potential has a very
physical
meaning, and is a consequence of the instabilities that
drive phase separation.

A rather clear understanding of the imaginary part and the
physics associated with it is obtained by a derivation of
the
effective potential within the Hamiltonian framework. The
Hamiltonian for a scalar field theory quantized in a volume
$\Omega$ is
\begin{equation}
H= \int_{\Omega} d^3x \left\{\frac{1}{2}\Pi^2(x)+\frac{1}{2}
(\vec{\nabla}\Phi(x))^2+V(\Phi)\right \} \label{hamiltonian}
\end{equation}

Since the effective potential is a function of the zero
momentum
component of the field, we separate the constant part (zero
momentum)
 ($\varphi$) of the field and its canonical momentum
\begin{eqnarray}
                         \Phi(x) & = & \varphi + \psi(x) \;
;  \; \; \varphi =
\frac{1}{\Omega}\int_{\Omega} d^3x \Phi(x) \; ; \; \;
\int_{\Omega}
d^3 x \psi(x) = 0 \label{fieldsplit} \\
                         \Pi(x)  & = & \frac{P}{\Omega} +
\pi(x) \; ; \; \;
 P= \int_{\Omega}
 d^3x \Pi(x) \; ; \; \; \int_{\Omega}d^3x \pi(x) =  0
\label{momentumsplit} \\
\left[ \Pi (x), \Phi (y) \right] & = & -i \hbar \delta^3 (x-
y) \; ; \; \;
 \left[ P,\varphi \right]  =  -i\hbar
\label{zeromodecommutator}
\end{eqnarray}

We have kept $(\hbar)$ in the above commutators to clarify
the quantum corrections.

Using that $\pi \; , \psi$, do not have zero momentum
component, the Hamiltonian becomes
\begin{equation}
H = \frac{P^2}{2 \Omega} + \Omega V(\varphi) +
\int_{\Omega}d^3x
\left\{\frac{1}{2}\pi^2(x) + \frac{1}{2}
(\vec{\nabla}\psi(x))^2+
\frac{1}{2}V''(\varphi) \psi^2(x)+\cdots \right \}
\end{equation}
where the ellipsis stand for higher order terms.

To this order
the Hamiltonian for the field $\psi(x)$ is quadratic, with a
mass term that depends on $\varphi$. In the representation
with
creation and annihilation operators for the harmonic
oscillators,
the Hamiltonian finally becomes (in the discrete momentum
representation)
\begin{eqnarray}
                H & = & \frac{P^2}{2\Omega}+ \Omega
V(\varphi) + \sum_k
(a^{\dagger}_k a_k + \frac{1}{2})\hbar \omega(k,\varphi)
\label{oscillators} \\
\omega(k,\varphi) & = & [k^2 + V''(\varphi)]^{\frac{1}{2}}
\label{frequency}
\end{eqnarray}
When all the oscillators are in their ground state, the zero
point energy for the oscillators is then recognized as the
first order quantum correction to the effective potential.
Taking
the large volume limit, the effective potential to this
order
is  then
\begin{equation}
V_{eff}(\varphi) = V(\varphi) +\frac{\hbar}{2} \int
\frac{d^3k}{(2\pi)^3}
\left[k^2+V''(\varphi)\right]^{\frac{1}{2}}
\label{effectivepotential}
\end{equation}
As usual renormalization is carried out in the standard
manner.
 From now on, we set $ \hbar=1$.
Consider for simplicity the case in which
\begin{equation}
V(\varphi) = -\frac{1}{2} \mu_0^2 \varphi^2+
\frac{\lambda_0}{4!}\varphi^4
\label{treepotential}
\end{equation}
then for all values of $\varphi$ such that $V''(\varphi)<0$,
the
effective potential acquires an imaginary part given by
\begin{eqnarray}
Im V_{eff}(\varphi)  & = & \pm
\frac{\hbar}{4\pi^2}\int_0^{k_{max}}k^2
\left[|V''(\varphi)|-k^2 \right]^{\frac{1}{2}}dk
\label{imaginarypart} \\
            (k_{max})^2 & = & |V''(\varphi)| \label{k_{max}}
\end{eqnarray}

This imaginary part is finite (independent of
renormalization),
and arises because for these values of $\varphi$ the modes
for which $k^2 < (k_{max})^2$ are unstable, the frequencies
of the oscillators in
(\ref{oscillators}) are imaginary. For these modes  the
potential is that of an {\it inverted oscillator}.

These inverted
oscillators for which $\omega^2(k,\varphi) <0$, do not have
stationary state solutions  and the imaginary part of the
zero point energy arises from
the {\it analytic continuation} of harmonic oscillator
wavefunctions
from positive  to negative $\omega^2$. The sign of the
imaginary part depends on the analytic continuation. The
quantum
mechanics of these inverted harmonic oscillators has been
thoroughly studied by Guth and Pi\cite{guthpi} and Weinberg
and  Wu\cite{weinbergwu}.

There being no stationary states associated with these
inverted
oscillators, an initial state must be specified.
If, for example, a gaussian wave packet is prepared
initially
 centered at the origin of these inverted oscillators, it
will spread  and the width
of the packet will increase exponentially in
time\cite{guthpi,weinbergwu}. This exponential growth, is
manifest in the equal time two point function\cite{guthpi,weinbergwu}
\[< \psi_k(t) \psi_{-k}(t) > \sim e^{2|\omega(k,\varphi)|t} \]
 This is the two-point Green's function computed
in this Gaussian state and measures the fluctuations of the
operator $\psi$. In particular, the growth of these unstable
modes is precisely the mechanism that Mazenko et. al.\cite{mazenko}
suggest.

One notes, however, that the two-point Green's function
calculated in the
{\it analytically continued state} is {\it time independent}
and  given by
\begin{equation}
< \psi_k(t) \psi_{-k}(t)> = \frac{1}{2\omega(k,\varphi)}
\label{continuedgreenfunction}
\end{equation}
This result is purely imaginary, but again this imaginary
part is
obtained after the analytic continuation of these Gaussian
states, and is actually concealing a {\bf time dependent
situation}.

Another crucial observation is that it is this two-point
function
evaluated in these Gaussian states
that gives the one-loop correction to the effective
potential.
Thus, it becomes clear, that the imaginary
part, resulting from an analytic continuation of the
unstable
modes is in fact hiding an unstable time evolution. At
finite
temperature, the imaginary part is signaling a non-
equilibrium situation.

This situation is well understood in statistical mechanics,
perhaps the earliest example is the Van der Walls equation
of state and its unphysical isotherms.
 In the unphysical region, the
situation must be studied out of equilibrium. In this region
there is coexistence of different phases that may not be
studied
within equilibrium statistical mechanics. An ad-hoc remedy
in this situation, is the Maxwell construction that replaces
the unphysical region of the isotherms by a straight line.
Its physical interpretation is that the non-equilibrium
state  may be
found, in the coexistence region, as a mixture of phases
with arbitrary
 concentrations of each phase. This situation is also
typical of binary mixtures in statistical
mechanics\cite{langer,guntonmiguel}.
The Maxwell constructed effective potential or free energy
is {\it irrelevant} for the dynamical non-equilibrium
description of the system.

The exponential growth of the unstable modes (inverted
oscillators) is signaling the growth of domains and the
onset of
the process that triggers the phase transition, i.e. phase
separation. This is very similar to the process of spinodal
decomposition in statistical
mechanics\cite{langer,guntonmiguel}.
An attempt to describe spinodal decomposition within the
context of field theory has been described by
Calzetta\cite{calzetta}.

Clearly the approximation of inverted oscillators, is crude
as it neglects non-linear effects.
The growth of these unstable modes
will eventually slow down when the non-linearities become
important, this is the process of coarsening.
As we will show
later, a clear understanding of the physics of coarsening
in the regime where the non-linearities become important,
 may require departing from perturbative treatments.

Usually in order to understand the dynamics of the scalar
field in this type of situations, the {\it static effective
potential} is used in the equations of motion,
resulting in a typical evolution equation (in flat
spacetime)
\begin{equation}
\frac{d^2 \varphi}{dt^2} +\frac{d
V_{eff}(\varphi)}{d\varphi} =0
\label{staticequationmotion}
\end{equation}

After the arguments presented above, it becomes clear that
this  equation is inappropriate for the study of the
dynamical
evolution. The imaginary part of $V_{eff}(\varphi)$ is
signaling
an unstability and the static effective potential
is not a suitable quantity to study the dynamics.
That is one must consider the effective action for {\it time
dependent
fields} and not merely the effective potential which holds
for constant
fields.

We turn to this study in the next section.

\section{\bf Non-Equilibrium time evolution}

As explained in the previous section, one must depart from
the usual
description in terms of the {\it static} effective
potential, and treat
the dynamics with the full time evolution. The time
evolution of the
system will be determined once the evolution Hamiltonian and
{\it the
initial state} are prescribed.

In order to understand the dynamics of the phase transition
and
the physics of the instabilities mentioned above, let us
consider
the situation in which for time $t<0$ the system is in {\it
equilibrium} at an initial temperature $T_i > T_c$ where
$T_c$ is the critical temperature.
At time $t = 0$ the system is  rapidly ``quenched''
to a final temperature below the critical temperature
$T_f < T_c$ and evolves thereafter out of equilibrium.

What we have in mind in this situation, is a cosmological
scenario with a period
of rapid inflation in which the temperature drops very fast
compared to typical relaxation times of the scalar field. In
particular this situation should correspond to the case
$h^{-1} \ll \tau$ with $h$ Hubble's constant and $\tau$ a
typical
relaxation time. At high temperatures and weakly coupled
theories
we would expect $\tau^{-1} \approx \lambda^2 T$ ($\lambda$
is
the coupling). When $T \approx T_c \propto \lambda ^{-
{1/2}}$
relaxation times become very
large and the dynamics of the long-wavelength modes (the
only
relevant modes for the phase transition) becomes
critically slowed down. Precisely because of this critical
slowing
down, we conjecture that an inflationary period at
temperatures near
the critical temperature, may be described in this
``quenched'' approximation.
Another situation that may be described by this
approximation is
that of a scalar field again at $T_i>T_c$ suddenly coupled
 to a ``heat bath'' at a
much lower temperature (below the transition temperature)
and evolving out of equilibrium. The ``heat bath'' may be
other
fields at a different temperature.  The influence of a
``heat bath'' in an inflationary universe, has been studied in the
linearized approximation by Cornwall and Bruinsma\cite{cornwall}.

 One then would expect that a ``quenching out of
equilibrium scenario'' may be an appropriate description
 near the critical temperature for these
situations. Certainly this is only a plausibility argument,
a deeper
understanding of the initial conditions must be pursued to
obtain a more precise knowledge of the cooling down process.

We do not envisage in this article to study the case of an
inflationary cosmology or the detailed dynamics of the
mechanism
that produces  the ``quenching'' below the critical
temperature, and departure from equilibrium.
We expect to report on these investigations in forthcoming
articles.

Here we just assume that such a mechanism takes place and
simplify the situation by introducing a Hamiltonian with a
{\it time dependent mass term} to describe this situation
\begin{eqnarray}
  H(t) & = & \int_{\Omega} d^3x \left\{
\frac{1}{2}\Pi^2(x)+\frac{1}{2}(\vec{\nabla}\Phi(x))^2+\frac
{1}{2}m^2(t)
\Phi^2(x)+\frac{\lambda}{4!}\Phi^4(x) \right \}
\label{timedepham} \\
m^2(t) & = & m^2 \Theta(-t) + (-\mu^2) \Theta(t)
\label{massoft}
\end{eqnarray}
where both $m^2$ and $\mu^2$ are positive. We assume that
for all times $t <0$ there is thermal equilibrium at
temperature $T_i$, and the system is described by the
density matrix
\begin{eqnarray}
\hat{\rho}_i  & = & e^{-\beta_i H_i} \label{initialdesmat}\\
          H_i & = & H(t<0) \label{initialham}
\end{eqnarray}
In the Schroedinger picture, the density matrix evolves in
time as
\begin{equation}
\hat{\rho(t)} = U(t)\hat{\rho}_iU^{-1}(t) \label{timedesmat}
\end{equation}
with $U(t)$ the time evolution operator.

An alternative and equally valid interpretation (and the one
that
we like best) is that the initial
condition  being considered here is that of a system in
equilibrium in the
symmetric phase, and evolved in time with a Hamiltonian that
allows for
broken symmetry ground states, i.e.  the Hamiltonian (\ref
{timedepham}, \ref{massoft}) for $t>0$.

The expectation
value of any operator is thus
\begin{equation}
< {\cal{O}} >(t) = Tr e^{-\beta_i H_i} U^{-1}(t)
{\cal{O}}U(t)/ Tr e^{-
\beta_i H_i}
\label{expecvalue}
\end{equation}
This expression may be written in a more illuminating form
by
choosing an arbitrary time $T <0$ for which
$U(T) = \exp[-iTH_i]$ then we may write
$\exp[-\beta_i H_i] = \exp[-iH_i(T-i\beta_i -T)] = U(T-
i\beta_i,T)$.
Inserting in the trace $U^{-1}(T)U(T)=1$,
commuting $U^{-1}(T)$ with $\hat{\rho}_i$ and using the
composition property of the evolution operator, we may write
(\ref{expecvalue}) as
\begin{equation}
< {\cal{O}}>(t) = Tr U(T-i\beta_i,t) {\cal{O}} U(t,T)/ Tr
U(T-i\beta_i,T) \label{trace}
\end{equation}
The numerator of the above expression has a simple meaning:
start at time $T<0$, evolve to time $t$, insert the operator
$\cal{O}$ and evolve backwards in time from $t$ to $T<0$,
and along the negative imaginary axis from $T$ to $T-
i\beta_i$.
This  operation is depicted in figure 1 (a).
The denominator, just evolves along the
negative imaginary axis from $T$ to $T-i\beta_i$. The
contour in
the numerator may be extended to an arbitrary large positive
time
$T'$ by inserting $U(t,T')U(T',t)=1$ to the left of
$\cal{O}$ in
(\ref{trace}), thus becoming
\begin{equation}
< {\cal{O}}>(t) = Tr U(T-
i\beta_i,T)U(T,T')U(T',t){\cal{O}}U(t,T)
/Tr U(T-i\beta_i,T)
\end{equation}
The numerator now represents the process of evolving from
$T<0$
to $t$, inserting the operator $\cal{O}$, evolving further
to
$T'$, and backwards from $T'$ to $T$ and down the negative
imaginary axis to $T-i\beta_i$. This process is depicted in
the
contour of figure 1(b). Eventually we take $T \rightarrow -
\infty
\; ; \; T' \rightarrow \infty$. It is straightforward to
generalize
to  real time correlation functions of Heisenberg picture
operators.

This formalism allows us also to study the general case in
which both the mass and the coupling depend on time, and
furthermore, by taking the zero temperature limit, we can
study the situation
in which a particular state is prepared at time $t=0$ and
evolved in time.

For example, by switching off the coupling for $t<0$
one is preparing a gaussian density matrix at $t=0$ or in
the zero temperature limit, a gaussian wavefunctional. This
density matrix or gaussian functional will then be evolved
in
time and in this time evolved state (or density matrix) we
compute expectation values of operators, or correlation
functions.
We then see that the above formalism permits us to study
these
situations in great generality.

As mentioned before, another point of view that one may take
on the
``quenching'' below the critical temperature, is that a
definite state or
density matrix describing the symmetric phase is prepared as
an initial condition  for $t<0$ and evolved in time  with
the
Hamiltonian that allows for broken symmetry states. One then
studies the dynamics of the phase transition, and
how the system evolves in time from the initially symmetric
state
towards the asymmetric states.

As usual, the insertion of an operator may be achieved by
inserting sources in the time evolution operators, defining
the generating functionals and eventually taking functional
derivatives with respect to these sources. Notice that we
have
three evolution operators, from $T$ to $T'$, from $T'$, back
to $T$ (inverse operator) and from $T$ to $T-i\beta_i$,
since
each of these operators has interactions and we want to use
perturbation theory and generate the diagrammatics from the
generating functionals, we use {\it three different
sources}.
A source $J^+$
for the evolution  $T \rightarrow T'$, $J^-$ for the branch
$T'\rightarrow T$ and finally $J^{\beta}$ for
$T \rightarrow T- i\beta_i$. The denominator may be obtained
from the numerator by
setting $J^+ = J^- =0$. Finally the generating functional
$Z[J^+, J^-, J^{\beta}]= Tr U(T-i\beta_i,T;J^{\beta})
U(T,T';J^-)U(T',T;J^+)$, may be written in term of path
integrals as (here we neglect the spatial arguments to avoid
cluttering of notation)
\begin{eqnarray}
Z[J^+,J^-,J^{\beta}] & = & \int D \Phi D \Phi_1 D \Phi_2
\int
{\cal{D}}\Phi^+ {\cal{D}}\Phi^-
{\cal{D}}\Phi^{\beta}e^{i\int_T^{T'}\left\{{\cal{L}}[\Phi^+,
J^+]-
{\cal{L}}[\Phi^-,J^-]\right\}}\times   \nonumber\\
                     &   &  e^{i\int_T^{T-
i\beta_i}{\cal{L}}[\Phi^{\beta}, J^{\beta}]}
\label{generfunc}
\end{eqnarray}
with the boundary conditions $\Phi^+(T)=\Phi^{\beta}(T-
i\beta_i)=\Phi \;
; \; \Phi^+(T')=\Phi^-(T')=\Phi_2 \; ; \; \Phi^-
(T)=\Phi^{\beta}(T)=
\Phi_1$.
As usual the path integrals over the quadratic forms may be
done and one obtains the final result for the partition
function
\begin{eqnarray}
Z[J^+,J^-, J^{\beta}] & = & e^{\left\{i\int_{T}^{T'}dt
\left[{\cal{L}}_{int}(-i\delta/\delta J^+)-
{\cal{L}}_{int}(i\delta/\delta
J^-)\right] \right \}}e^{\left\{i\int_{T}^{T-
i\beta_i}dt{\cal{L}}_{int}(-i\delta/\delta J^{\beta})
\right\}}
\times \nonumber \\
                      &   & e^{\left\{\frac{i}{2}\int_c
dt_1\int_c dt_2 J_c(t_1)J_c(t_2)G_c (t_1,t_2) \right\}}
 \label{partitionfunction}
\end{eqnarray}
Where $J_c $ are the currents defined on the contour
of figure (1,b) $J^{\pm}\; ,\; J^\beta$\cite{niemisemenoff}
and $G_c$ is the Green's function on the
contour (see below), and again the spatial arguments have
been suppressed.

In the two contour integrals (on $t_1 ; \; \; t_2$) in
(\ref{partitionfunction}) there are altogether nine terms,
corresponding to the combination of currents in each of the
three branches.
However, in the limit $T \rightarrow -\infty$, the
contributions
arising from the terms in which one current is on the $(+)$
or $(-)$ branch and another on the imaginary time segment
(from T to $T-i\beta_i$), go
to zero when computing correlation functions in which the
external
legs are at finite {\it real time}. For this {\it real time
correlation functions} there is no contribution from the
$J^{\beta}$ terms, that cancel between numerator and
denominator, and the information on finite
temperature is encoded in the boundary conditions on the
Green's
functions (see below). Then for the calculation of finite
{\it real time} correlation functions the generating
functional
simplifies to\cite{calzetta,calzettahu}
\begin{eqnarray}
Z[J^+,J^-] & = & e^{\left\{i\int_{T}^{T'}dt\left[
{\cal{L}}_{int}(-i\delta/\delta J^+)-
{\cal{L}}_{int}(i\delta/\delta
J^-)\right] \right \}} \times \nonumber \\
           &   & e^{\left\{\frac{i}{2}\int_T^{T'}
dt_1\int_T^{T'} dt_2 J_a(t_1)J_b(t_2)G_{ab} (t_1,t_2)
\right\}}
 \label{generatingfunction}
\end{eqnarray}
with $a,b = +,-$.

This formulation in terms of time evolution along a contour
in complex time has been used many times in non-equilibrium
statistical mechanics. To our knowledge the first to use
this formulation were Schwinger\cite{schwinger} and
Keldysh\cite{keldysh} (see also Mills\cite{mills}).
 There are many articles in the literature
using these techniques to study time dependent problems,
some of the more clear articles are by Jordan\cite{jordan},
Niemi and Semenoff\cite{niemisemenoff},
Landsman and van Weert\cite{landsman},Semenoff and
Weiss\cite{semenoffweiss}, Kobes and
Kowalski\cite{kobeskowalski}, Calzetta and
Hu\cite{calzettahu},Paz\cite{paz} and references therein.

At first sight one seems to have complicated the
situation enormously by doubling the number of fields.
However, this doubling is a natural consequence of dealing
with a time evolution of a {\it density matrix} and in
general with probabilities, instead of amplitudes. Rather
than computing in-out amplitudes, we are here computing
expectation
values or correlation functions in the time evolved in-state
or density matrix\cite{schwinger,jordan}.

  The Green's functions that enter in the integrals along
the  contours in (\ref{partitionfunction},
\ref{generatingfunction})
  are given by (see above references)
\begin{eqnarray}
G^{++}(t_1,t_2)  & = & G^{>}(t_1,t_2)\Theta(t_1 - t_2) +
G^{<}(t_1,t_2)\Theta(t_2-t_1) \label{timeordered}\\
G^{--}(t_1,t_2)  & = & G^{>}(t_1,t_2)\Theta(t_2-t_1) +
G^{<}(t_1,t_2)\Theta(t_1-t_2) \label{antitimeordered} \\
G^{+-}(t_1,t_2)  & = & -G^{<}(t_1,t_2) \label{plusminus}\\
G^{-+}(t_1,t_2)  & = & -G^{>}(t_1,t_2) = -G^{<}(t_2,t_1)
\label{minusplus}\\
G^{<}(T,t_2)     & = & G^{>}(T-i\beta_i,t_2)
\label{periodicity}
\end{eqnarray}

As usual $G^{<},G^{>}$ are homogeneous solutions of the
quadratic form with appropriate boundary conditions. We will
construct them explicitly later.
The condition (\ref{periodicity}) is recognized as the
periodicity condition in imaginary time (KMS
condition)\cite{kadanoffbaym}. It is straightforward to
show,  using the
above Green's functions, that $Z[J,J]=1$ as it must.

Although most of the details presented above on the non-
equilibrium
formalism are available in the literature, we included them
here
for self-consistency and with the intention to clarify some
issues that are usually glossed over in most treatments.

 We are now in condition to obtain the evolution equations
for the average of the scalar field in the case when the
potential is
suddenly changed, to account for a sudden change in
temperature
from above to below the critical
temperature as described by the model Hamiltonian
(\ref{timedepham}) with (\ref{massoft}).
 For this purpose we use the tadpole method\cite{sweinberg},
and write
\begin{equation}
\Phi^{\pm}(\vec{x},t) = \phi(t) + \Psi^{\pm}(\vec{x},t)
\label{expecvaluepsi}
\end{equation}
Where, again, the $\pm$ refer to the branches for forward
and backward time propagation. The reason for shifting both
($\pm$) fields by the {\it same} classical configuration, is
that $\phi$ enters in the time evolution operator as a
background c-number variable, and time evolution forward and
backwards are now considered in this background.

The  evolution equations are obtained with the
tadpole method by expanding the Lagrangian around $\phi(t)$
and considering the {\it linear}, cubic, quartic, and higher
order terms in $\Psi^{\pm}$ as perturbations and requiring
that
\[ <\Psi^{\pm}(\vec{x},t) >= 0 .\]

It is a straightforward exercise to see that this is
equivalent to
extremizing
the one-loop effective action in which the determinant (in
the logdet)
incorporates the boundary condition of equilibrium at time
$t<0$ at
the initial temperature.

To one loop we find the  equation of motion
\begin{equation}
\frac{d^2\phi(t)}{dt^2}+m^2(t)\phi(t)+\frac{\lambda}{6}
\phi^3(t)+ \frac{\lambda}{2}\phi(t)\int \frac{d^3
k}{(2\pi)^3}(-i) G_k(t,t) = 0 \label{eqofmotion}
\end{equation}
with $G_k(t,t)=G_k^{<}(t,t)=G_k^{>}(t,t)$ is the spatial
Fourier transform of the equal-time Green's function.

At this point, we would like to remind the reader that

\[(-iG_k(t,t)) =<\Psi^+_{\vec{k}}(t) \Psi^+_{-
\vec{k}}(t)>\]

is a {\it positive definite quantity} (because the field
$\Psi$ is real) and as we argued before
(and will be seen explicitly shortly) this Green's function
grows in time because of the instabilities associated with
the
phase transition and domain growth\cite{guthpi,weinbergwu}.

These Green's functions are constructed out of the
homogeneous solutions to the operator of quadratic
fluctuations
\begin{eqnarray}
\left[\frac{d^2}{dt^2} + \vec{k}^2 +
M^2(t)\right]{\cal{U}}_k^{\pm} & = & 0 \label{homogeneous}\\
 M^2(t)  =  (m^2+\frac{\lambda}{2}
\phi^2_i)\Theta(-t)            &   &
+(-\mu^2+\frac{\lambda}{2}\phi^2(t))\Theta(t)
\label{bigmassoft}
\end{eqnarray}

The boundary conditions on the homogeneous solutions are
\begin{eqnarray}
{\cal{U}}_k^{\pm}(t<0) & = & e^{\mp i
\omega_{<}(k)t}\label{boundaryconditions} \\
\omega_{<}(k)          & = &
\left[\vec{k}^2+m^2+\frac{\lambda}{2}\phi^2_i\right]^{\frac{
1}{2}}\label{omegaminus}
\end{eqnarray}
where $\phi_i$ is the value of the classical field at time
$t<0$ and is the initial boundary condition on the equation
of motion. Truly speaking, starting in a fully symmetric
phase
will force $\phi_i =0$, and the time evolution will maintain
this value, therefore we admit a small explicit symmetry
breaking field in the initial density matrix to allow for a
small $\phi_i$. The introduction of this initial condition
seems artificial since we are studying the situation of
cooling down from the symmetric phase.

 However, we recognize that  the phase transition from the
symmetric phase
 occurs via formation of domains (in the case of a discrete
symmetry)
inside which the order parameter acquires non-zero values.
The domains
will have the same probability for either value of the field
and the
volume average of the field will remain zero. These domains
will grow in time, this is the phenomenon of phase
separation and spinodal decomposition
familiar in condensed matter physics. Our evolution
equations presumably
will apply to the coarse grained average of the scalar field
inside each
of these domains. This average will only depend on time.
Thus, we interpret
$\varphi_i$ as corresponding to the coarse grained average
of the field in each of these domains.
The question of initial conditions on the scalar field is
also present (but usually overlooked) in the slow-rollover
scenarios but as we will see later, it plays a fundamental
role in the description of the evolution.

The identification of the initial value $\varphi_i$ with the
average of the field in each domain is certainly a
plausibility argument to
justify an initially small asymmetry in the scalar field
which is necesary
for the further evolution of the field, and is consistent
with the usual
assumption within the slow rollover scenario.

We are currently studying  the dynamics of the phase
transition from the symmetric phase by looking at the
composite operator
$\Phi^2(\vec{x},t)$ which measures the fluctuations, and
will report on our studies in a forthcoming
article\cite{anupam}. An alternative approach using this
composite operator has  also been proposed by
Lawrie\cite{lawrie}.

 The boundary conditions on the mode functions
${\cal{U}}_k^{\pm}(t)$ correspond to  ``vacuum'' boundary
conditions of positive and negative frequency modes
(particles and antiparticles) for $t<0$.

Finite temperature enters through the periodicity conditions
(\ref{periodicity}) and the Green's functions are
\begin{eqnarray}
G^{>}_k(t,t') & = & \frac{i}{2\omega_<(k)} \frac{1}{1-e^{-
\beta_i \omega_<(k)}}\left[{\cal{U}}_k^+(t) {\cal{U}}_k^-
(t')+ e^{-\beta_i\omega_<(k)} {\cal{U}}_k^-(t)
{\cal{U}}_k^+(t') \right] \label{finalgreenfunc} \\
G^{<}_k(t,t') & = & G^{>}(t',t)
\end{eqnarray}

Summarizing, the effective equations of motion to one loop
that determine the time evolution of the scalar field are
\begin{eqnarray}
\frac{d^2\phi(t)}{dt^2}+m^2(t)\phi(t) & + &
\frac{\lambda}{6}
\phi^3(t) + \nonumber \\
                                      &   &
               \frac{\lambda}{2}\phi(t) \int \frac{d^3
k}{(2\pi)^3} \frac{{\cal{U}}_k^+(t) {\cal{U}}_k^-(t)
}{2\omega_<(k)} \mbox{coth}\left[\frac{\beta_i
\omega_<(k)}{2}\right]
 =  0 \label{finaleqofmotion1} \\
\left[\frac{d^2}{dt^2} + \vec{k}^2 +
M^2(t)\right]{\cal{U}}_k^{\pm}
          & = & 0 \label{finaleqofmotion2}
\end{eqnarray}
with (\ref{bigmassoft}) , (\ref{boundaryconditions}).

This set of equations is too complicated to attempt an
analytic
solution, we will study this system numerically shortly.

However, before studying numerically these equations, one
recognizes that
there are several features of this set of equations
that reveal the basic physical aspects of the dynamics of
the
scalar field.

{\bf i)}: The effective evolution equations are {\bf real}.
The mode functions ${\cal{U}}_k^{\pm}(t)$ are complex
conjugate of  each other
as may be seen from the time reversal symmetry of the
equations,
and the boundary conditions (\ref{boundaryconditions}).
This  situation must be contrasted with the expression for
the
effective potential for the {\it analytically continued
modes}.

{\bf ii)}: Consider the situation in which the initial
configuration of the classical field is near the origin
$\phi_i \approx 0$,
for $t>0$, the modes for which $\vec{k}^2 < (k_{max})^2 \; ;
\;
(k_{max})^2 = \mu^2-\frac{\lambda}{2}\phi_i^2$ are {\it
unstable}.

In particular, for early times ($t>0$), when $\phi_i \approx
0$,
these unstable modes behave approximately as
\begin{eqnarray}
{\cal{U}}_k^+(t) & = & A_k e^{W_k t}+B_k e^{-W_k t}
\label{unstable1}\\
{\cal{U}}_k^-(t) & = & ({\cal{U}}_k^+(t))^{*}
\label{unstable2}\\
A_k              & = & \frac{1}{2}\left[1-
i\frac{\omega_{<}(k)}
{W_k} \right] \; ; \; B_k = A_k^* \label{abofk} \\
W_k              & = & \left[\mu^2-\frac{\lambda}{2}\phi_i^2
- \vec{k}^2 \right]^{\frac{1}{2}} \label{bigomega}
\end{eqnarray}
Then the early time behavior of $(-iG_k(t,t))$ is given by
\begin{equation}
(-iG_k(t,t)) \approx \frac{1}{2\omega_<(k)}
\left[1+\frac{\mu^2+m^2}
{\mu^2-\frac{\lambda}{2} \phi_i^2-k^2}[cosh(2W_kt)-1]\right]
coth[\beta_i\omega_<(k)/2]
\label{earlytimegreen}
\end{equation}
This early time behavior coincides with the Green's function
of Guth and
Pi\cite{guthpi} and Weinberg and Wu\cite{weinbergwu} for the
inverted harmonic oscillators when our initial
state (density matrix) is taken into account.

Our evolution equations, however, permit us to go beyond the
early
time behavior and to incorporate the non-linearities that
will eventually shut off the instabilities.

These early-stage instabilities and subsequent growth of
fluctuations
and correlations, are the hallmark of the process of phase
separation,
and precisely the instabilities that trigger the phase
transition.

It is
clear from the above equations of evolution, that the
description in terms
of inverted oscillators will only be valid at very early
times.

At early times, the {\it stable} modes for which
$\vec{k}^2>(k_{max})^2$
are obtained
from (\ref{unstable1}) , (\ref{unstable2}) , (\ref{abofk})
by the  analytic continuation
$ W_k \rightarrow -i \omega_{>}(k)= \left[ \vec{k}^2-\mu^2+
\frac{\lambda}{2}\phi_i^2 \right]^{\frac{1}{2}} $.

For $t<0$, ${\cal{U}}_k^+(t){\cal{U}}_k^-(t)=1$ and one
obtains
the usual result for the evolution equation
\[ \frac{d^2 \phi(t)}{dt^2}+\frac{dV_{eff}(\phi)}{d\phi} =0
\]
with $V_{eff}(\phi)$ the finite temperature effective
potential
but for $t<0$ there are no unstable modes.

It becomes clear, however, that for $t>0$ there are no {\it
static}
solutions to the evolution equations for $\phi(t) \neq 0$.

{\bf iii)}{\bf Coarsening}: as the classical expectation
value
$\phi(t)$ ``rolls down'' the potential hill, $\phi^2(t)$
increases and $(k_{max}(t))^2= \mu^2-
\frac{\lambda}{2}\phi^2(t)$
{\it decreases}, and only the very
long-wavelength modes remain unstable, until for a
particular
time  $t_s\; ; \; (k_{max}(t_s))^2=0$. This occurs when
$\phi^2(t_s) = 2\mu^2/\lambda$, this is the inflexion point
of the
tree level potential. In statistical mechanics this point is
known
as the ``classical spinodal point'' and $t_s$ as the
``spinodal time''\cite{langer,guntonmiguel}. When the
classical
field reaches the spinodal point, all instabilities shut-
off.
 From this point on, the dynamics is oscillatory and this
period is
identified with the ``reheating'' stage in cosmological
scenarios\cite{linde,brandenberger}.

It is
clear from the above equations of evolution, that the
description in terms
of inverted oscillators will only be valid at small positive
times, as eventually
the unstable growth will shut-off.

The value of the spinodal time depends on the
initial conditions of $\phi(t)$. If the initial value
$\phi_i$ is
very near the classical spinodal point, $t_s$ will be
relatively
small and there will not be enough time for the unstable
modes
to grow too much.
In this case, the one-loop corrections for small coupling
constant
will remain perturbatively small.
On the other hand, however, if $\phi_i \approx 0$, and the
initial velocity is small, it will take a very long time to
reach the
classical spinodal point. In this case the unstable modes
{\bf may grow
dramatically making the one-loop corrections non-negligible
even for
small coupling}. These initial conditions of small initial
field and
velocity are precisely the ``slow rollover'' conditions that
are of interest in cosmological scenarios of ``new
inflation''.

 {\bf Renormalization}:

 As argued above, for $t>0$, there are no static solutions
to
 the equation of motion for the scalar field. The mode
 functions ${\cal{U}}^{\pm}_k(t)$ depend {\it implicitly} on
the field
 $\phi(t)$. As in the usual situation, one expects
ultraviolet
 divergences in the one-loop correction. It is not clear
from the
 equation of motion for the scalar field whether these
divergences may
 be absorbed in a redefinition of the mass and coupling
constant, or
 cancelled by {\it local} counterterms. Since we want to
study the
 time evolution and be able to extract meaningful
information, we
 must first understand the renormalization aspects of the
effective
 equation of motion.

 The ultraviolet divergences must be absorbed in $m^2(t)$
and
 $\lambda$, whose coefficients in the equation of motion are
 $\phi(t)$ and $\phi^3(t)$ respectively, i.e. {\it time
 dependent}. Since as mentioned previously, for $t<0$ the
 situation corresponds to the usual case of the static
effective
 potential, renormalization proceeds in the standard manner.
 However for $t>0$ the situation is different and to
understand
 it we need the large $k$ behavior of the mode functions. We
study
 the short wavelength behavior by a WKB-type analysis: we
define
 $\epsilon = 1/k$ and divide the equation for the mode
functions by
 $k^2$ thus obtaining the equation
 \begin{equation}
 \left[\epsilon^2 \frac{d^2}{dt^2}+1+\epsilon^2M^2(t)\right]
 {\cal{U}}_k^{\pm}(t) = 0 \label{redequation}
 \end{equation}
 with the boundary conditions (\ref{boundaryconditions}).
 Let us define the functions ${\cal{V}}_k^{\pm}(t)$
 as the two linearly independent solutions for $t>0$ to the
 equation(\ref{redequation}) with the boundary conditions
 ${\cal{V}}_k^{\pm}(0^+)=1$. The mode functions
 ${\cal{U}}_k^{\pm}(t)$ solutions
 to (\ref{redequation}) with the boundary conditions
 (\ref{boundaryconditions}) are written as
 \begin{eqnarray}
  {\cal{U}}_k^{+}(t) & = & c_k {\cal{V}}_k^{+}(t)+d_k
  {\cal{V}}_k^{-}(t) \\
  {\cal{U}}_k^{-}(t) & = & ({\cal{U}}_k^{+}(t))^*
  \end{eqnarray}
  the coefficients $c_k$, $d_k$ are obtained from the
matching conditions at $t=0$.
 We propose a WKB ansatz for the mode functions
${\cal{V}}_k^{\pm}(t)$
 \begin{eqnarray}
 {\cal{V}}_k^{\pm}(t) & = & e^{i\frac{S^{\pm}(t)}{\epsilon}}
 \label{wkb} \\
 S^{\pm}(t)           & = & \sum_{n=0}\epsilon^n
S_n^{\pm}(t)
 \label{soft}
 \end{eqnarray}
 inserting this ansatz in (\ref{redequation}) and comparing
 powers of $\epsilon$ we find the asymptotic behavior for
large k
 to be
 \begin{equation}
  {\cal{V}}_k^{\pm}(t) = e^{\mp i(kt+\frac{1}{2k}\int_0^t
  M^2(t')dt')}
  \left[1-\frac{1}{4k^2}(M^2(t)-M^2(0^+))\right]+\cdots
  \label{wkbfunctions}
 \end{equation}
 The leading behavior for large k of the coefficients is
found to be
 \begin{eqnarray}
 c_k & = & \frac{1}{2}\left[1 + \frac{\omega_<(k)}
 {k+\frac{M^2(0^+)}{2k}}\right]+\cdots \label{cofk} \\
 d_k & = & \frac{1}{2}\left[1 - \frac{\omega_<(k)}
 {k+\frac{M^2(0^+)}{2k}}\right]+\cdots \label{dofk} \\
 \end{eqnarray}

 Inserting these results for the large k behavior in the
one-loop
 contribution, it is straightforward to find that the
divergent
 terms are independent of temperature and we obtain
 \begin{equation}
 \int\frac{d^3k}{(2\pi)^3} \frac{{\cal{U}}_k^{+}(t)
 {\cal{U}}_k^{-}(t)}{2\omega_<(k)}
 \mbox{coth}\left[\frac{\beta_i \omega_<(k)}{2}\right]
 = \frac{1}{4\pi^2}\Lambda^2 -\frac{1}{4\pi^2}(-\mu^2+
 \frac{\lambda}{2}\phi^2(t))\ln(\frac{\Lambda}{\kappa})+
 \mbox{finite}
 \label{divergences}
 \end{equation}
 where $\Lambda$ is an upper momentum cutoff, and $\kappa$ a
 renormalization scale and the finite part is  time,
temperature
 and $\kappa$ dependent.

It is clear that these divergences may be cancelled by local
counterterms of the usual form, where the mass counterterm
depends (locally) on time and changes suddenly at $t=0$. The
coupling constant is renormalized in the usual manner.

 It is important to point out that the integral for the one-
loop
 correction may be split into the contribution from the
unstable
 modes $k^2 < (k_{max}(t))^2$ and that of the stable modes
 $k^2 > (k_{max}(t))^2$. It is only the latter that requires
 renormalization and where the divergences reside.
 The contribution from the
 unstable modes is {\it finite} and does not require
 renormalizations. The renormalization of the effective
action has been done in an alternative manner using
dimensional renormalization by Avan and de Vega\cite{veg}.

 \section{\bf Analysis of the Evolution}

 As mentioned previously within the context of coarsening,
when
 the initial value of the scalar field $\phi_i \approx 0$,
and
 the initial temporal derivative is small, the scalar field
slowly
 rolls down the potential hill. But during the time while
the scalar
 field remains smaller than the ``spinodal'' value, the
unstable
 modes grow and the one-loop contribution grows
consequently. For a
 ``slow rollover'' condition, the field remains very small
($\phi^2(t)
 \ll 2\mu^2/\lambda$) for a long
 time, and during this time the unstable modes grow
exponentially.
 The stable modes, on the other hand give an oscillatory
contribution
 which is bound in time, and for weak coupling remains
perturbatively
 small at all times.

 Then for a ``slow rollover'' situation and for
 weak coupling, only the unstable modes will yield to an
important
 contribution to the one-loop correction. Thus, in the
evolution equation
 for the scalar field, we will keep only the integral over
the {\it
 unstable modes} in the one loop correction.

Phenomenologically the coupling constant in these models is
bound by the spectrum of density fluctuations to be within
the range $\lambda_R \approx 10^{-12} - 10^{-
14}$\cite{linde}. The the stable modes will
\underline{always} give a \underline{perturbative}
contribution, whereas the unstable modes grow exponentially
in time thus raising the possibility of giving a non-
negligible contribution.

 With the purpose of numerical analysis of the effective
equations
 of motion, it proves convenient to introduce the following
 dimensionless variables
 \begin{eqnarray}
 \tau      & = & \mu_R t \; ; \;  q   =  k/\mu_R \\
 \eta^2(t) & = & \frac{\lambda_R}{6\mu_R^2}\phi^2(t) \; ; \;
L^2 =
 \frac{m_R^2 + \frac{1}{2}\lambda_R \phi_i^2}{\mu^2_R}
 \end{eqnarray}
 and to account for the change from the initial temperature
to the final
 temperature ($T_i>T_c \; ; \; T_f<T_c$) we
parametrize\cite{note}
 \begin{eqnarray}
 m^2   & = & \mu_R(0)\left[\frac{T_i^2}{T_c^2} -1 \right]
\label{abovecrit} \\
 \mu_R & = & \mu_R(0)\left[1-\frac{T_f^2}{T_c^2} \right]
\label{belowcrit}
 \end{eqnarray}
 where the subscripts (R) stand for  renormalized
quantities, and
 $-\mu_R(0)$ is the renormalized zero temperature ``negative
mass squared''
 and $T^2_c = 24 \mu^2_R(0)/\lambda_R$.  Furthermore,
because $(k_{max}(t))^2
 \leq \mu^2_R$ and $T_i > T_c$, for the unstable modes $T_i
\gg (k_{max}(t))$
 and we can take the high temperature limit
$\mbox{coth}[\beta_i
 \omega_<(k)/2] \approx 2 T_i/\omega_<(k)$.  Finally the
effective
 equations of evolution for $t > 0$, keeping in the one-loop
contribution only the unstable
 modes  as explained above ($q^2 <(q_{max}(\tau))^2$)
become, after using $\omega_<^2 = \mu_R^2(q^2+L^2)$
 \begin{eqnarray}
 \frac{d^2}{d\tau^2}\eta(\tau)-\eta(\tau)& + & \eta^3(\tau)+
\nonumber \\
                                         &   & g\eta(\tau)
\int_0^{(q_{max}(\tau))}q^2\frac{{\cal{U}}^+_q(\tau){\cal{U}}^-_q(\tau)}
  {q^2+L^2}dq = 0 \label{eqofmotunst}\\
  \left[\frac{d^2}{d\tau^2}+ q^2-
(q_{max}(t))^2\right]{\cal{U}}^{\pm}_q(\tau)
                                         & = & 0
\label{unstamod} \\
                       (q_{max}(\tau))^2 & = & 1-3\eta^2(\tau)
  \label{qmaxoft} \\
                                      g  & = & \frac{\sqrt{6\lambda_R}}
  {2\pi^2}\frac{T_i}{T_c \left[1-\frac{T^2_f}{T^2_c}\right]}
\label{coupling}
  \end{eqnarray}

For $T_i \geq T_c$ and $T_f \ll T_c$ the coupling (\ref{coupling}) is bound
within the range $g \approx 10^{-7}-10^{-8}$.
 The dependence of the coupling with the temperature
reflects the
 fact that at higher temperatures the fluctuations are
enhanced.
 It is then clear, that the contribution from the stable
modes
 is {\it always perturbatively small}, and only the unstable
modes may
 introduce important corrections if they are allowed to grow
for a long  time.

 From (\ref{eqofmotunst}) we see that the quantum
corrections act as a
 {\it positive dynamical renormalization} of the ``negative
mass'' term
 that drives  the rolling down dynamics. It is then clear,
that the
 quantum corrections tend to {\it slow down the evolution}.

In particular,
 if the initial value $\eta(0)$ is very small, the unstable
modes grow for
 a long time before $\eta(\tau)$ reaches the spinodal point
$\eta(\tau_s)
 = 1/\sqrt{3}$ at which point the instabilities shut off. If
this is the
 case, the quantum corrections introduce substantial
modifications to the
 classical equations of motion, thus becoming non-
perturbative. If
 $\eta(0)$ is closer to the classical spinodal point, the
unstable modes
 do not have time to grow dramatically and the quantum
corrections are
 perturbatively small.

 Thus we conclude that the initial conditions on the field
determine
 whether or not the quantum corrections are perturbatively
small.

 Although the system of equations (\ref{eqofmotunst},
\ref{unstamod}),
 are coupled, non-linear and integro-differential, they may
be integrated
 numerically. Figures (2,3,4) depict the solutions for the
classical\cite{noteclas}
 (solid lines) and quantum (dashed lines) evolution (a),
 the quantum correction (b), i.e. the fourth term in
(\ref{eqofmotunst})
 {\it including the coupling} $g$, the classical (solid
lines) and
 quantum (dashed lines) velocities
$\frac{d\eta(\tau)}{d\tau}$ (c)
 and $(q_{max}(\tau))^2$ (d). For the numerical integration,
we
have
 chosen $L^2 =1$, the results are only weakly dependent on
$L$, and
 taken $g=10^{-7}$, we have varied the initial condition
$\eta(0)$ but
 used $\frac{d\eta(\tau)}{d\tau}|_{\tau=0} = 0$.

 We recall, from a previous discussion that $\eta(\tau)$ should be
 identified with the average of the field within a domain. We are
 considering the situation in which this average is very small, according
 to the usual slow-rollover hypothesis, and for  which the instabilities
 are stronger.

 In figure (2.a)
 $\eta(0) = 2.3 \times 10^{-5}$ we begin to see that the
quantum  corrections become important at $t \approx 10/\mu_R$ and
slow down the dynamics. By the time that the {\it classical}
evolution  reaches the minimum of the classical potential at $\eta
=1$, the quantum evolution has just reached the classical spinodal $\eta =
1/\sqrt{3}$.
 We see in figure (2.b) that the quantum correction becomes
large enough
 to change the sign of the ``mass term'', the field
continues its
 evolution towards larger values, however, because the
velocity is
 different from zero, attaining a maximum (figure (2.c))
around the
 time when the quantum correction attains its maximum. As
$\eta$ gets
 closer to the classical spinodal point, the unstability
shuts off as is
 clearly seen in figure (2.d) and the quantum correction
arising from the
 unstable modes become perturbatively small. From the
spinodal point
 onwards, the field evolves towards the minimum and begins
to oscillate
 around it, the quantum correction will be perturbatively
small, as all
 the instabilities had shut-off. Higher order corrections,
will introduce
 a damping term as quanta may decay in elementary
excitations of the true
 vacuum.

 Figures (3.a-d) show a more marked behavior for $\eta(0)=
2.27 \times
 10^{-5}\; ; \; \frac{d\eta (0)}{d\tau}=0$, notice that the
classical
 evolution of the field has reached beyond the minimum of
the potential
 at the time when the quantum evolution has just reached the
classical
 spinodal point. Figure (3.b) shows that the quantum
correction becomes
 larger than $1$, and dramatically slows down the evolution,
again
 because the velocity is different from zero (figure (3.c))
the field
 continues to grow. The velocity reaches a maximum and
begins to drop.
 Once the field reaches the spinodal again the instabilities
shut-off
 (figures (3.b,d)) and from this point the field will
continue to evolve
 towards the minimum of the potential but the quantum
corrections will
 be perturbatively small.

 Figures (4.a-d) show a dramatic behavior for $\eta(0) =
2.258 \times
 10^{-5} \; ; \; \frac{d\eta (0)}{d\tau} = 0$. The unstable
modes have
 enough time to grow so dramatically that the quantum
correction (figure
 (4.b)) becomes extremely large $\gg 1$ (figure(4.b)),
overwhelming the
 ``negative mass'' term near the origin.
 The dynamical time dependent potential, now becomes
 {\it a minimum} at the origin and the quantum evolution
begins to
 {\it oscillate} near $\eta = 0$. The contribution of the
unstable modes
 has become {\it non-perturbative}, and certainly our one-
loop
 approximation breaks down.

 As the initial value of the field gets closer to zero, the
unstable modes
 grow for a very long time. At this point, we realize,
however, that this
 picture cannot be complete. To see this more clearly, consider the case
 in which the
initial state
 or density matrix corresponds exactly to the symmetric
case. $\eta =0$ is
 necessarily, by symmetry, a fixed point of the equations of
motion.
 Beginning from the symmetric state, the field will {\it
always remain} at
 the origin and though there will be strong quantum and
thermal fluctuations,
 these are symmetric and will sample field configurations
with opposite
 values of the field with equal probability.

 In this situation, and according to the picture presented
above, one would
 then expect that the unstable modes will grow indefinetely
because the
 scalar field does not roll down and will never reach the
classical spinodal
 point thus shutting-off the instabilities. What is missing
in this picture
 and the resulting equations of motion is a self-consistent
treatment of the
 unstable fluctuations, that must necessarily go beyond one-
loop. A more
 sophisticated and clearly non-perturbative scheme must be
invoked that will
 incorporate coarsening, that is the shift with time of the
unstable modes
 towards longer wavelength and the eventual shutting off of
the instabilities.
 We are currently\cite{anupam} exploring a Hartree
approximation that will
 incorporate self-consistently these features. Another possible approach
 would be a variational treatment as advocated in references (18-19) or
 as proposed by Lawrie\cite{lawrie}.

 {\bf Conclusions:}

 We have studied the effective equations of motion for the spatially
 independent average  of a scalar field evolving out of equilibrium
after a second
 order phase transition. After pointing out the severe
shortcomings in the
 usual description in terms of the effective potential, we
considered the
 situation in which a scalar
 field theory is suddenly cooled below the critical
temperature from an
 initial state at a temperature higher than the critical
temperature. We use
 the tools of statistical mechanics out of equilibrium to
study
 the {\it real time} dynamics during a  ``slow rollover
stage''.
 The effective non-equilibrium equations of motion are
studied
 both analytically and numerically to one-loop level.
 We find
 that the unstable growth of long wavelength fluctuations,
that trigger
 the process of phase separation is responsible of a very
marked behavior
 in the time evolution of the scalar field. Even for very
weak couplings,
 consistent with the bounds from density fluctuations,
 for the case of ``slow rollover'' initial conditions, the
time evolution
 is dramatically {\it slowed down} as a consequence of this
unstable growth
 which signal the onset of the phase transition.
 When the scalar field is very  close to the origin (at the
local maximum
 of the potential) the unstable modes can grow for a long
time and the
 effect of the quantum (and thermal) corrections become very large and
 eventually non-perturbative.
 We give a qualitative and  quantitative description  of the
coarsening
 process and the eventual shut-off of the instabilities.

 We argue that a comprehensive treatment and understanding
of the late stage
 dynamics of the phase transition, and time evolution of the
scalar field,
 involves a {\it non-perturbative} treatment out of equilibrium.
This treatment
 must incorporate not only the growth but also the
coarsening features  when
 the initial state is symmetric and the initial value of the
scalar field is
 zero, thus remaining zero during the time evolution.

 This is  the next step in a consistent analysis of the
dynamics, work
 on this problem is currently underway. Clearly, all this
must be extended to
 the case of inflationary cosmology, in particular a
description of the
 dynamics out of equilibrium in a de Sitter background is
the next stage.
 In this case the physics of coarsening will be complicated
by the red-shift
 of wavevectors and more modes enter the unstable regime,
clearly a deeper
 understanding of these instabilities is necessary.

 However we believe that the results presented in this
article
 provide a novel insight into the long standing problem of
the dynamics of
 a phase transition and the ensuing evolution out of
equilibrium when quantum and thermal fluctuations are taken into
account. It also  points out that this problem must be studied as
 a fully time dependent process and that one must abandon
the usual treatment in terms of the effective potential.

The potential consequences for inflationary scenarios are obvious, the
instabilities and growth of long-wavelength fluctuations enhance the
quantum and thermal corrections that become large. As a consequence,
the ``slow-rollover'' stage is delayed, the scalar field remains near the
``false vacuum'' for a longer period of time giving rise to a longer
inflationary stage and a delayed completion of the phase transition.

\acknowledgments

The authors would like to deeply thank R. Willey for insightful remarks,
lengthy conversations and discussions, and for paving the way on some
sticking points.
D.B. would like to thank R. Holman, I. Lawrie, A. Singh and D.-S. Lee for
fruitful discussions and comments. He would also like to thank the LPTHE
at the Universite de Pierre et Marie Curie for warm hospitality during part of
this work and the N.S.F. for support through the Grant: PHY-8921311.
H. de Vega, would like to thank the Physics Department at the
U. of Pittsburgh for hospitality during part of this work.

\newpage

{\bf Figure Captions}:

\vspace{1mm}

\underline{Figure 1(a):} Contour of evolution in complex
time plane, the
cross denotes insertion of an operator.

\underline{Figure 1(b):} Final contour of evolution,
eventually $T'
\rightarrow \infty$ , $T \rightarrow -\infty$.

\underline{Figure 2(a):} $\eta$ vs $\tau$ (notation in the
text) for
$g= 10^{-7}$, $\eta(0) = 2.3 \times 10^{-5}$, $\eta'(0) =
0$, $L=1$.
The solid line is the classical evolution, the dashed line
is the evolution from the one-loop corrected equation (\ref{eqofmotunst}).
\underline{Figure 2(b):} One-loop contribution including the
coupling $g$ for
the values of the parameters used in Figure 2(a).

\underline{Figure 2(c):} Velocity $\frac{\partial
\eta}{\partial \tau}$, same
values and conventions as in figure 2(a).

\underline{Figure 2(d):} $(q_{max}(\tau))^2$ vs $\tau$ for
the same values as in figure 2(a).

\underline{Figure 3(a):} $\eta$ vs $\tau$ with $g=10^{-7}$
$\eta(0) = 2.27 \times 10^{-5}$, $\eta'(0) = 0$, $L=1$. Same
conventions for solid and dashed lines as in figure 2(a).

\underline{Figure 3(b):} One-loop contribution
to the equations of motion including the
coupling $g$ for the values of the parameters used in Figure 3(a).

\underline{Figure 3(c):} Velocity $\frac{\partial
\eta}{\partial \tau}$, same
values and conventions as in figure 3(a).

\underline{Figure 3(d):} $(q_{max}(\tau))^2$ vs $\tau$ for
the same values as in figure 3(a).

\underline{Figure 4(a):} $\eta$ vs $\tau$ with $g=10^{-7}$
$\eta(0) = 2.258 \times 10^{-5}$, $\eta'(0) = 0$, $L=1$.
Same conventions for  solid and dashed lines as in figure 2(a).

\underline{Figure 4(b):} One-loop contribution including the
coupling $g$ for the values of the parameters used in Figure 4(a).

\underline{Figure 4(c):} Velocity $\frac{\partial
\eta}{\partial \tau}$, same values and conventions as in figure 4(a).

\underline{Figure 4(d):} $(q_{max}(\tau))^2$ vs $\tau$ for
the  same values as in  figure 4(a).


\begin{references}
\bibitem{note} This may be done formally by generalizing the method proposed
by S. Weinberg\cite{sweinberg} and introduce a time dependent counterterm
in the original Hamiltonian of the form $\frac{\lambda}{24}\left(T_i^2
\theta(-t)+T_f^2 \theta(t)\right)$ and cancelling this counterterm against
the one loop contribution for $t<0$.
\bibitem{noteclas} Where by classical solution we mean the solution to
$\frac{d^2 \eta_{cl}}{d \tau^2}-\eta_{cl}+\eta_{cl}^3 =0$, with the same
boundary conditions. Its solution with $\frac{d \eta}{d\tau}|_0 = 0$ and
$\eta(0) \ll 1$ is a periodic elliptic function.
\end{references}
\end{document}